\documentclass{ws-ijmpa}

\usepackage{amsmath}
\def\fsl#1{\setbox0=\hbox{$#1$}                 
   \dimen0=\wd0                                 
   \setbox1=\hbox{/} \dimen1=\wd1               
   \ifdim\dimen0>\dimen1                        
      \rlap{\hbox to \dimen0{\hfil/\hfil}}      
      #1                                        
   \else                                        
      \rlap{\hbox to \dimen1{\hfil$#1$\hfil}}   
      /                                         
   \fi}                                         %

\begin{document}

\markboth{Mei Huang}{Understanding magnetic instability}

\catchline{}{}{}{}{}

\title{Understanding magnetic instability in gapless superconductors}

\author{Mei Huang}
\address{Physics Department, 
     University of Tokyo, 
     Hongo, Bunkyo-ku, Tokyo 113-0033, Japan}

\maketitle

\vspace{-5.6cm}

\hfill {\small TKYNT-05/21 }

\vspace{5.6cm}


\begin{abstract}

Magnetic instability in gapless superconductors still remains as a puzzle. In this
article, we point out that the instability might be caused by using BCS theory in mean-field 
approximation, where the phase fluctuation has been neglected. The mean-field BCS theory 
describes very well the strongly coherent or rigid superconducting state.  
With the increase of mismatch between the Fermi surfaces of pairing 
fermions, the phase fluctuation plays more and more important role, and ``soften" the
superconductor. The strong phase fluctuation will eventually 
quantum disorder the superconducting state, and turn  the system into
a phase-decoherent pseudogap state.

\end{abstract}



\section{Introduction}


In recent years, it attracts lots of interest on the charge neutral color superconducting 
state with a moderate mismatch between the Fermi surfaces of the pairing quarks. 
It was found that homogeneous neutral cold-dense quark 
matter can be in the gapless 2SC (g2SC) phase \cite{g2SC} or gapless CFL (gCFL) 
phase \cite{gCFL}, depending on the flavor structure of the system.  
Unexpectedly, it was found that in the g2SC phase, the Meissner screening masses for 
five gluons corresponding to broken generators of $SU(3)_c$ become imaginary, which 
indicates a type of chromomagnetic instability in the g2SC phase \cite{chromo-ins-g2SC}.
The calculations in the gCFL phase show the same type of chromomagnetic instability 
\cite{chromo-ins-gCFL}.  Remembering the discovery of superfluidity density instability 
\cite{Wu-Yip} in the gapless interior-gap state 
\cite{Liu-Wilczek}, it seems that the instability is a inherent property of gapless phases. 

The chromomagnetic instability or anti-Meissner effect in the gapless color superconducting
phase remains as a puzzle. The anti-Meissner effect is contrary to our knowledge on superconductor, 
whose most distinguishing feature is the Meissner effect, i.e., the superconductor expels the magnetic 
field \cite{Meissner-cond}. The Meissner effect can be understood using the standard Anderson-Higgs 
mechanism. In ideal color superconducting phases, e.g., in the ideal 2SC and CFL phases, the 
gauge bosons connected with the broken generators obtain masses, which indicates the Meissner 
screening effect \cite{Meissner}.  

It was suggested in Ref. \cite{LOFF-Ren} that the chromomagnetic instability in the gapless superconducting phases indicates the formation of the LOFF(Larkin-Ovchinnikov-Fudde-Ferrell) 
state \cite{loff-orig,LOFF}. Latter on, it was found that the LOFF-like state can be driven through 
different ways, e.g., by spontaneous generation of baryon current \cite{huang-current}, by 
Goldstone boson supercurrent \cite{hong-current},  or by gluon 
condensate \cite{miransky-gluon-cond}. 

In this article, we offer a new point of view on the magnetic instability problem in the gapless
superconductors.  We point out that the instability might be caused by using BCS theory in mean-field 
approximation, where the phase fluctuation has been neglected. 

\section{How a superconductor will be destroyed?}

We start from some general arguments. It is noticed that the magnetic instability is induced
by increasing the mismatch between the Fermi surfaces of the Cooper pairing.
A superconductor will be eventually destroyed and goes to the normal Fermi liquid state,
so one natural question is: how an ideal superconductor will be destroyed by increasing 
mismatch? 

To answer the question how a superconductor will be destroyed, one has to firstly understand 
what is a superconductor. The superconducting phase is characterized by the order parameter 
$\Delta(x)$, which is a complex scalar field and has the form of
e.g., for electrical superconductor, 
$\Delta (x) = |\Delta| e^{i\varphi (x)}$, with $|\Delta| $ the amplitude and $\varphi $ the phase
of the gap order parameter.
1) The superconducting phase is charaterized by the nonzero vacuum expectation value, i.e., 
$<\Delta>\neq 0$, which means the amplitude of the gap is finite, and the phase coherence 
is also established;  2) If the amplitude is still finite, while the phase coherence is lost, this phase
is in a phase decoherent  pseudogap state characterized by $|\Delta|\neq 0$, 
but  $<\Delta> =|\Delta| <e^{i\varphi(x)}> = 0$; 
3) The normal state is characterized by $|\Delta|=0$.

There are two ways to destroy a superconductor: 1) by driving the amplitude of the
oder parameter to zero. This way is BCS-like, because it mimics the behavior of a 
conventional superconductor at finite temperature, the gap amplitude monotonously 
drops to zero with the increase of temperature;
2) Another way is non-BCS like, but Berezinskii-Kosterlitz-Thouless (BKT)-like \cite{BKT},  
even if the amplitude of the order parameter is large and finite, superconductivity will be 
lost with the destruction of phase coherence, e.g.  the phase transition from the $d-$wave
superconductor to the pseudogap state in high temperature superconductors 
\cite{Emery-Kivelson}. 

\section{The role of phase fluctuation}

Since all the gapless superconducting systems exhibit magnetic instability, we
use a minimal model for the gapless superconducting phase as introduced in 
Ref. \cite{hong-current}.  There are two different flavors, up and down, in the 
system. The pairing force is $SU(2)_c$ color interaction, and the Cooper pair is  
color-singlet. The system is an electric superconductor and exhibits the essential 
features of gapless superconductivity.  The system is described by a Lagrangian density
${\cal L}_q={\bar q}\left( 
i \fsl{D}  + {\hat \mu} \gamma_0 \right) q
+G_\Delta [\,(\bar{q}^C i\varepsilon\epsilon\gamma_5 q)
(\bar{q} i\varepsilon\epsilon\gamma_5 q^C)\,] $, with $D_\mu \equiv \partial_\mu - ie A_\mu$,
here  $A_{\mu}$ is the electromagnetic gauge field. ${\hat \mu}$ is chemical potential in flavor
space, with $\mu_u={\bar \mu} -\delta\mu$ and $\mu_d= {\bar \mu}+\delta\mu$, $\delta\mu$ is
the mismatch between the fermi surfaces of the pairing fermions. $\varepsilon$ and $\epsilon$ are the 
antisymmetric tensors in the $SU(2)$ flavor and $SU(2)$ color spaces, respectively. The 
bosonized Lagrangian takes the form of
$ {\cal L}_q^b = \bar{q}(i\fsl{D}+\hat \mu \gamma^0)q
 -\frac{1}{2}\Delta[i\bar{q}\varepsilon\epsilon \gamma_5 q^C]
 -\frac{1}{2}[i\bar{q}^C\varepsilon\epsilon\gamma_5 q]\Delta^{*}
 - \frac{|\Delta|^2}{4G_\Delta}$, where $\Delta = |\Delta| e^{i \varphi}$ is the 
order parameter for superconductor. 

In conventional BCS superconductor, where the superfluid density is large,
the phase fluctuation is absent. However, with the increase of $\delta\mu$, the superfluid 
density decreases, and the phase fluctuation plays more and more important role. The role of  phase 
fluctuation has been totally neglected in all the previous papers on gapless superconductors. 
In the following, we consider the contribution from the phase fluctuation.  

In order to couple the phase fluctuation to the quasiparticles, one has to 
isolate the uncertain charge carried by quasiparticles $q$ as in high temperature
superconductors. 
We preform the Franz-Tesanovic (FT) singular gauge transformation as introduced in Ref. \cite{FT}, 
${\bar \psi}_u = e^{i\varphi_u} {\bar q}_u$, and ${\bar \psi}_d = e^{i\varphi_d} {\bar q}_d$, 
with $\varphi_u+\varphi_d=\varphi$. Note that this gauge transformation
defines a new set of charge neutral fermions $\psi$. The topological defects are indicated by 
$\triangledown \times \triangledown\varphi_{u(d)} = 2\pi{\hat z}\sum_i Q_i \delta({\vec r} -{\vec r}_i^{u(d)})$
with $Q_i$ the topological charge of $i-$th Abrikosov-Nielsen-Olesen string \cite{ANO-string}
and ${\vec r}_i^{u(d)}$ its position.  
The full Lagrangian takes the form ${\cal L}_{\psi}
=  {\cal L}_{\psi}^{qp}+{\cal L}_{\psi}^{a,v}$, with
${\cal L}_{\psi}^{qp}={\bar \psi}\left( i \fsl{{\tilde D}} + {\hat \mu} \gamma_0 \right) \psi
-\frac{1}{2}|\Delta|[i\bar{\psi}\varepsilon\epsilon \gamma_5 {\psi}^C]
 -\frac{1}{2}[i\bar{\psi}^C\varepsilon\epsilon\gamma_5 {\psi}]|\Delta|
 - \frac{|\Delta|^2}{4G_\Delta} $, and ${\cal L}_{\psi}^{a,v}=\frac{1}{4\pi^2 |\Phi|^2}(V_{\mu\nu}+A_{\mu\nu})$,
 where ${\tilde D}= (\partial_{\mu} + i 2a_{\mu} ) + i ( v_{\mu} -  e A_{\mu})$.
The two emergent vector fields $v_{\mu} = \frac{1}{2}\partial_{\mu}\varphi$ is a Doppler gauge field 
or superfluid field,  and $a_{\mu}=\frac{1}{2}(\partial_{\mu}\varphi_u- \partial_{\mu}\varphi_d)$ is a Berry or topological gauge field.  $V_{\mu\nu}, A_{\mu\nu}$ are strength tensor for $v_{\mu},a_{\mu}$. A dual 
 filed $\Phi$ has been introduced to quantum disorder the superconducting phase, with
 $<\Phi>=0$ indicating the superconducting phase and $<\Phi>\neq0$ indicating the phase
 decoherent pseudogap state. For a more general description for the dual disorder field \cite{dual}, 
 please see Ref. \cite{FT}. 
 
After including the phase fluctuation, the expected phase diagram is: when the mismatch is
small, the system is in conventional BCS superconducting phase, where the superfluid density
is large and the system is strongly coherent and rigid, thus the phase fluctuation can be neglected. 
When the mismatch increases, the phase fluctuation starts
to play some role, and "soften" the superconductor. 
At some critical mismatch, the phase fluctuation
becomes very strong, and destroy the long-range phase coherence, the system is in a phase decoherent
pseudogap state where the amplitude of the order paramter is still finite, the low-energy degrees of
freedom in this state are gapless quasiparticles and massless emergent vector fields. 
With further increase of mismatch, the amplitude of the gap is driven to zero, and system goes to a 
normal state. It is noticed that the existence of the phase decoherent pseudogap state is dependent on
the assumption that the phase fluctuation is stronger than the amplitude fluctuation. 


\section{Conclusion}

In this article, we offer a new point of view on the magnetic instability problem in the gapless
superconductors.  We point out that the instability is caused by using BCS theory in mean-field 
approximation, where the phase fluctuation has been totally neglected.  
With the increase of mismatch, the phase 
fluctuation plays more and more important role, which ``softens" the superconductor. The strong phase 
fluctuation will eventually destroy the long-range phase coherence, and quantum disorder the 
superconducting state to a phase decoherent pseudogap state.

\vskip 2mm
{\bf Acknowledgements}
The author thanks T. Hatsuda, Z. Tesanovic, Z.Y. Weng  for valuable discussions. 
The work is supported by the Japan Society for the Promotion of Science (JSPS) fellowship program.

\end{document}